\documentclass{elsart}
\usepackage{epsfig,t1enc}
\begin{document}
\begin{frontmatter}
\title{The far-infrared absorption of a periodic 2DEG\\
       in the transition regime between \\ weak and strong modulation}
\author{Vidar Gudmundsson},
\author{Ingibj{\"o}rg Magn{\'u}sd{\'o}ttir}, \\ and
\author{Sigurdur I.\ Erlingsson}
\address{Science Institute, University of Iceland,\\
        Dunhaga 3, IS-107 Reykjavik, Iceland.\\
        E-mail: vidar@raunvis.hi.is,\\
        Fax: 00354-552 8911.}
\begin{abstract}
      We study the optical absorption of arrays of quantum 
      dots and antidots in a perpendicular homogeneous 
      magnetic field. The electronic system is described
      quantum mechanically using a Hartree approximation 
      for the mutual Coulomb interaction of the electrons.
      The evolution of the absorption is traced
      from the homogeneous to the strongly modulated case
      identifying the ensuing collective modes, the 
      magnetoplasmons, and their correlations with 
      inherent length scales of the system.
\end{abstract}
\thanks{This research was supported in part by the Icelandic Natural Science
        Foundation, the University of Iceland Research Fund, and the 
        Graduierten\-kolleg {\lq}Nano\-struktur\-ierter Fest\-körper', DFG.} 
\end{frontmatter}

%---------------------- Introduction --------------------------
%
Technically it is possible to tune a gate-modulated lateral 
superlattice from the case of quantum dots to antidots. 
In a short period structure the consistent inclusion of the
Coulomb interaction is essential in order to correctly 
model the ground state and the far-infrared (FIR) absorption
of the two-dimensional electron gas (2DEG).      

%
%---------------------- Model ---------------------------------
%
The square array of quantum dots or antidots is represented by the
periodic potential 
\begin{equation}
      V_{\mathrm{per}}({\bf r})=V\left\{ \sin\left(\frac{\pi x}{L}\right)
      \sin\left(\frac{\pi y}{L}\right) \right\} ^2 \quad ,
\label{Vper}
\end{equation}
where $L$ is the periodic length of the array. 
The ground-state properties
of the interacting 2DEG in a perpendicular homogeneous magnetic field
and the periodic potential are
calculated within the Hartree approximation for the Coulomb
interacting electrons at a finite
temperature \cite{Gudmundsson95:16744}. 
The FIR absorption of the system is calculated within
the time-dependent Hartree approximation perturbing the 2DEG
with an incident electric field 
\begin{equation}
      {\bf E}_{\mathrm{ext}}({\bf r},t)=-iE_0\frac{{\bf k}}
      {|{\bf k}|}\exp{\left\{ i{\bf k}\cdot{\bf r}
      -i\omega t\right\} } 
\label{Eext}
\end{equation}
with finite, but small wavevector
$|(k_xL,0)|=0.2$. The power absorption is found as the Joule heating
of the self-consistent time-dependent electric field composed of the
external field and the induced
field \cite{Gudmundsson96:5223R}. 
%
%---------------------- Results -------------------------------
%

\begin{figure}[tbh]
      \vspace*{-5.4cm}
      \centering\epsfig{figure=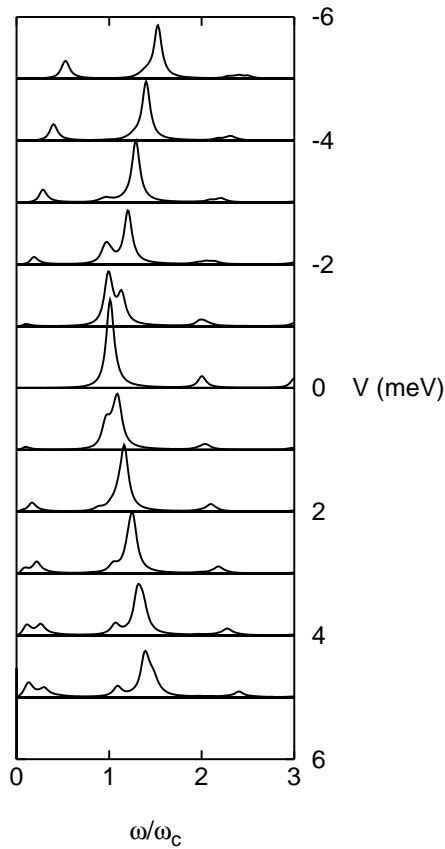,width=7.5cm}
      \vspace*{-1.0cm} 
      \caption{The FIR absorption as a function of the scaled
               frequency $\omega /\omega_c$ and the modulation $V$.
               $L=100$ nm, $B=1.24$ T, $N_s=0.5$, $l=23$ nm, and $T=1$ K.}
\label{P_V}
\end{figure}

We use GaAs parameters, $m^*=0.067m_0$, $\kappa =12.4$, and
assume $L=100$ nm and $T=1$ K. The absorption is calculated for
the magnetic field $B=1.24$ T, leading to the cyclotron energy
$\hbar\omega_c=2.14$ meV and the magnetic length $l=23$ nm.  
In figure \ref{P_V} the absorption is presented as a 
function of the external frequency $\omega$ for several values
of the modulation $V$.
The electron density is low, on the average one electron per two unit
cells of the array, $N_s=0.5$. This is possible since we are dealing with 
extended Bloch-like states. The model can thus not describe what
happens in the tight-binding limit.   
In the unmodulated case, $V=0$, three Bernstein peaks are 
visible
\cite{Gudmundsson96:5223R,Bernstein58:10,Gudmundsson17744:95,Horing76:216},
(The one with the highest frequency is located just at the right
edge of the figure).
The generalized Kohn theorem predicting one magnetoplasmon peak 
is broken by the small wavevector of the incident electric field.
The general structure of the peaks is otherwise not modified
by the small nonzero wavevector.   
Slight modulation, positive or negative, causes the main Bernstein
peak to broaden giving it two maxima reflecting the van Hove 
singularities of the active band, (the band the elctrons are 
excited into). The main peak is blue-shifted 
with increasing modulation as the energy separation of the 
bands is determined by $(\hbar\omega_c)^2$ and $V^2$ for low
electron density. 

\begin{figure}[tbh]
      %\vspace*{-5.4cm}
      \centering\epsfig{figure=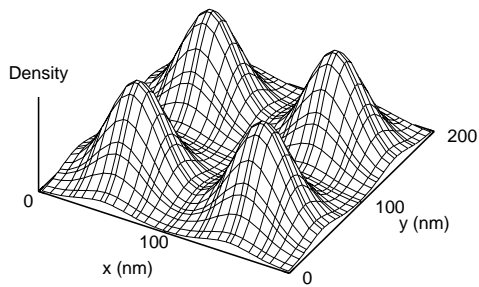,width=7.5cm}
      %\vspace*{-1.0cm} 
      \caption{The electron density of a quantum dot array
               in arbitrary units
               in four unit cells. Two electrons occupy
               the shown area, $N_s=0.5$.
               $L=100$ nm, $B=0.41$ T, $l=40$ nm, and $T=1$ K.}
\label{Dens}
\end{figure}

For the evolving dot array, $V<0$, a simple
two peak structure emerges as the electrons get increasingly localized
in the dot minima. The {\lq}confining' potential 
the electrons see approaches    
slowly the parabolic form necessary for the Kohn's center-of-mass
modes as can by verified by the time-dependent induced density.
If the magnetic field is lower than here, and $l$ thus larger,
the confinement will be poorer and more effects of the band structure
due to the periodic array are visible. In other words, the coupling
of individual dots is stronger. Figure \ref{Dens} shows the electron
density in the case of the lower magnetic field $B=0.41$ T. 

In the limit of an antidot array,
$V>0$, we also obtain essentially the expected two peak structure.
The antidot array supports a richer variety of collective modes
that depend sensitively on the relation of $l$ to $L$
\cite{Vasiliadou95:R8658,Mikhailov96:10335}. The induced density
can be used to identify the lower peaks with skipping orbits around
individual antidots, the upper peaks with oscillation of the density
between four antidots. Still higher peaks are caused by perturbed linear
waves travelling along the array in the direction of the incident
radiation but acquiring a wavelength $L$, rather than the one 
imposed on the system by the external electric field.       
 
To which extent the single dots in an array are coupled can be
fine tuned by the magnetic field $B$, the electron density $N_s/L^2$, 
and the modulation $V$ assuming that $L$ is constant. Even 
for the strongest modulation chosen here, $V=-5$ meV, band
structure or coupling effects can be brought into play by
reducing the magnetic field.    

The evolution of the Berstein modes can here be traced from the
unmodulated homogeneous 2DEG to the cases of arrays of dots or
antidots. Their existence has also been verified 
experimentally in isolated wires and dots, and in 
self-consistent models of single
dots and wires \cite{Gudmundsson17744:95}.
 
%
%------------------------ Biblio --------------------------------
%
\bibliographystyle{prsty}
\bibliography{../../BIBtex/mod_qd}
%----------------------------------------------------------------
\end{document}